\documentclass[letterpaper,twocolumn,10pt]{article}
\usepackage{usenix, epsfig, endnotes}
\usepackage{amsmath, amsthm, amssymb, natbib}
\usepackage{url, breakurl, hyperref}
\usepackage{multirow}
\usepackage{verbatim}

\newcommand{\tooEz}{$\sim$2EZ\/}

\begin{document}

\date{\today}

%make title bold and 14 pt font (Latex default is non-bold, 16 pt)
\title{\Large \bf  Limiting Risk by Turning Manifest Phantoms into Evil Zombies}

%for single author (just remove % characters)
\author{%I.M.~Anonymous
{\rm Jorge H.~Ba\~{n}uelos}\\
{\rm Philip B.~Stark}\\
Department of Statistics\\University of California, Berkeley\\[1ex]
{\rm \today}
% copy the following lines to add more authors
% \and
% {\rm Name}\\
%Name Institution
} % end author

\maketitle

% Use the following at camera-ready time to suppress page numbers.
% Comment it out when you first submit the paper for review.
% \thispagestyle{empty}

\abstract{
Drawing a random sample of ballots to conduct a risk-limiting audit
generally requires knowing how the ballots cast in an election
are organized into groups, for instance, how many containers of ballots there
are in all and how many ballots are in each container.
A list of the ballot group identifiers along with number of ballots in each group
is called a {\em ballot manifest\/}.
What if the ballot manifest is not accurate?
Surprisingly, even if ballots are {\em known\/} to be missing
from the manifest, it is not necessary to make
worst-case assumptions about those ballots---for instance, to adjust the margin by the
number of missing ballots---to ensure that the audit remains conservative.
Rather, it suffices to make worst-case assumptions about the individual randomly 
selected ballots that the audit cannot find. 
This observation provides a simple modification to some risk-limiting audit procedures that
makes them automatically become more conservative if the ballot manifest has errors.
The modification---{\em phantoms to evil zombies\/} (\tooEz)---requires
only an upper bound on the total number of ballots
cast.
\tooEz~makes the audit $P$-value stochastically larger than it would
be had the manifest been accurate, automatically requiring more than enough
ballots to be audited to offset the manifest errors.
This ensures that the true risk limit remains smaller than the nominal risk limit.
On the other hand, if the manifest is in fact accurate and the upper bound on the
total number of ballots equals the total according to the manifest,
\tooEz~has no effect at all on the number of ballots audited nor on the true risk limit.
}

\section{Introduction}
Election results can be wrong for a variety of reasons, including programming errors,
hardware malfunctions,
voter errors, pollworker errors, ballot definition errors, lapses in protocol, accidents, and deliberate fraud.
%%\citep{bonner,frisina,garber,mebane,lindemanStark12,starkWagner12}.
Properly designed and executed post-election vote tabulation audits can catch and
correct errors in election outcomes.

The current gold standard for post-election audits is a risk-limiting audit
\citep{bestPractices08,stark08a,stark09d,stark09b,stark10d,checkowayEtal10,benalohEtal11,lindemanStark12}.
An audit is risk-limiting if it has a pre-specified minimum probability of
progressing to a full hand count of the audit trail if
the reported outcome is wrong, no matter why it is wrong; the full hand count then
corrects the reported outcome.
Of course, for a full hand count of the audit trail to reveal the correct outcome,
the audit trail must be sufficiently intact and accurate to reflect the correct outcome;
a compliance audit \citep{benalohEtal11,lindemanStark12,starkWagner12}
can assess the evidence that this is true.
And there must be an audit trail in the first place.

Compliance audits (e.g., \citet{benalohEtal11,lindemanStark12,starkWagner12})
collect evidence about the accuracy and integrity of the audit trail to determine
whether, as actually used in the election, the voting system
was strongly software independent~\citep{rivestWack06,rivest08}.
Absent strong evidence that the audit trail is sufficiently accurate, a risk-limiting
audit is mere theater, because
risk-limiting audits rely on the accuracy of the audit trail.
The combination of a strongly software-independent voting system,
a compliance audit to assess whether there is strong affirmative evidence
that the audit trail is adequately accurate, and a risk-limiting audit of the
audit trail, comprises a resilient canvass framework~\citep{benalohEtal11,starkWagner12}.
A resilient canvass framework tends to recover from errors.
If it reports a winner, there is high statistical confidence that the reported winner is
the real winner.

Risk-limiting audits rely not only on the accuracy of the audit trail itself:
They also rely on the accuracy of information {\em about\/} the audit trail,
typically exported from the voting system or compiled by the local election official
during the canvass.
In particular, risk-limiting audits generally take these as given: the total number of ballots,
the organization of those ballots into groups, such as containers of ballots,
and the number of ballots in each group.
(Some auditing methods also require knowing the reported votes in each group.)
A listing of the groups of ballots and the number of ballots in each group is called
a {\em ballot manifest\/}~\citep{lindemanStark12}.
Generally, designing and carrying out the audit so that each ballot
has the correct probability of being selected involves the ballot manifest.

While elections officials generally are quite good at keeping track of paper,
nothing is perfect:
There are benign exceptions\footnote{%
In our experience, a few ballots from one precinct are sometimes misfiled with
ballots from another precinct, causing the manifest to be off for both
precincts.
}
and monstrous exceptions.\footnote{%
E.g.,
\burl{http://www.stamfordadvocate.com/news/article/Bridgeport-vote-recount-shows-widespread-876032.php\#page-1},
last accessed 8~May 2012;\\
\burl{http://www.wired.com/threatlevel/2008/10/florida-countys},
last accessed 6~May 2012;\\
\burl{http://www.tulsaworld.com/news/article.aspx?subjectid=11&articleid=20120501_16_A1_CUTLIN910968},
last accessed 7~May 2012.
}
Hence, ballot manifests are sometimes inaccurate.
The audit could in principle re-count all the ballots in every group and construct
its own ballot manifest from scratch, but statistics might let us
spend less and still maintain the risk-limiting property of the audit.
How?

\tooEz~is a very minor change to two risk-limiting audit procedures---ballot-level comparison
audits and ballot-polling audits~\cite{lindemanStark12}---that automatically
makes the procedures {\em more\/}
conservative if the ballot manifest has errors
\tooEz~ensures that the nominal $P$-value (loosely speaking, the chance of
observing sample results that favor the reported winner as much as they do, if
the reported winner were not the true winner) is stochastically larger\footnote{%
   The random variable $X$ is stochastically larger than the random variable $Y$ if, for every
   threshold $t$, $\Pr (X \ge t) \ge \Pr (Y \ge t)$.
}
than it would be if the ballot manifest had been known perfectly---so ballots had been
drawn uniformly and independently.
This gives the audit an even larger chance of leading to a full hand count
if the outcome is wrong and the manifest is wrong
than it would have had if the manifest had been right,
keeping the risk limit below the nominal risk limit.

We address ballot-level comparison
audits (e.g., \citet{stark10d,benalohEtal11,lindemanStark12}
and ballot-polling audits (e.g., \citet{lindemanStark12,lindemanEtal12}),
both of which seek to draw individual ballots with equal probability.
However, the same general approach can be adapted to any sampling scheme
so far proposed for risk-limiting audits of plurality contests, vote-for-$k$ contests,
or contests requiring a majority or super-majority.
We do not explore extending the method to ranked-choice voting (RCV)
or instant-runoff voting (IRV).
We do not claim that \tooEz~is in any sense optimal.
Rather, we show that there is a very easy way to automatically address some kinds of errors
in the manifest, reducing the set of things the compliance audit needs to check
to ensure that the resulting overall process comprises a resilient canvass framework
\citep{benalohEtal11,starkWagner12}.

\tooEz~needs an upper bound on the true total number of ballots, which could be
provided by the compliance audit or the local election official.
(The number of registered voters eligible to vote in the contest under audit is a
weak upper bound; ballot accounting and counting pollbook signatures give sharper bounds.)
The compliance audit also needs to ensure that no ballots were lost, added, altered, or substituted---or
to ensure that the number of such ballots is so small that it cannot alter the outcome.
And the risk-limiting
audit needs to take that number into account, treating the uncertainty in the most
pessimistic way~\citep{starkWagner12}.
If there is {\em no\/} upper bound on the true total number of ballots, or if the upper bound is
too weak, the method we propose here is not helpful---but we doubt any method could be.

The next few sections of this brief paper present special cases and the general case
of \tooEz.
We start with the case that the true total number of ballots, $N_O$, is known.
(In this notation, $O$ stands for ``oracle'': We imagine that an oracle tells us
the true total number of ballots.)
We then examine the case that $N_O$ is not known, but an upper-bound $N_U \ge N_O$
is known.

\section{Ballot-level comparison audits and ballot-polling audits}
\citet{stark10d,benalohEtal11,lindemanStark12} discuss a simple
method for ballot-level comparison audits
based on the maximum across-contest pairwise overstatement of margins (MACRO)
\citep{stark08d,stark09d}
and a conservative simplification of the Kaplan-Markov $P$-value~\citep{stark09b}.
The method compares a human interpretation of the votes on ballots selected at random (uniformly,
with replacement)
to the voting system's interpretation of the same ballots,
continuing to draw and examine ballots until there is strong evidence that the winners
reported by the voting system are the true winners.
The strength of evidence is measured by the $P$-value; smaller $P$-values are stronger evidence
that the reported outcome is correct.
If the $P$-value falls below the risk limit, the audit stops; otherwise, after some number of draws,
the auditors conduct a full hand count, which reveals the correct outcome and overrides the
voting-system outcome if they disagree.

A {\em pairwise margin\/} is the difference between the number of votes for any
winner in a contest and any loser in that contest.
If the contest allows voters to select $k$ of $n$ candidates, then there would be $k(n-k)$
pairwise margins in that contest: Each of the $k$ winners can be paired with each of the
$n-k$ losers.
In a typical plurality contest, $k = 1$; a city council or school board contest might have $k = 3$,
for instance.

Suppose that the human interpretation of the ballots does not match the machine interpretation.
Imagine correcting the machine interpretation of the ballot to match the human interpretation.
Some pairwise margins might not change; others might increase; and others might decrease.
The magnitude of the change to any margin is at most 2~votes.

If correcting the machine interpretation would increase any
pairwise margin in any contest under audit by 2~votes, the ballot has a 2-vote overstatement.
If the ballot does not have a 2-vote overstatement, but correcting the machine
interpretation would increase any
pairwise margin in any contest under audit by 1~vote, the ballot has a 1-vote overstatement.
If the ballot has neither a 2-vote overstatement nor a 1-vote overstatement, but
correcting the machine interpretation would leave any pairwise margin in any contest under audit
unchanged,  the ballot has a 0-vote overstatement.
If the ballot does not have a 2-vote overstatement, a 1-vote overstatement, or a 0-vote overstatement,
but correcting the machine interpretation would increase any pairwise margin in any contest under audit
by exactly one vote, $-1$-vote overstatement.
Otherwise, the ballot has a $-2$-vote overstatement.

For instance, if there is any contest under audit in which the machine reported a vote for the winner
but a manual interpretation reveals a vote for one of the losers in that contest,
the ballot has a 2-vote overstatement, no matter whether or how any other (winner, loser) pairs are
affected.
If the ballot does not have a 2-vote overstatement, but
there is a contest under audit in which the machine reported an undervote but a manual interpretation
reveals a vote for one of the losers of that contest, the ballot has a 1-vote overstatement.

In this ballot-level comparison audit, the $P$-value depends on the number of
ballots in the sample with overstatements of 2, 1, 0, -1, or -2~votes.
Ballots that show a maximum overstatement of 2~votes increase the $P$-value most;
ballots with a maximum overstatement of 1~vote also increase the $P$-value, but by less.
Ballots that show a maximum overstatement of 0, -1, or -2 votes decrease the $P$-value.
Hence, substituting 2 for the actual maximum overstatement of any ballot in the sample
cannot decrease the $P$-value.

Ballot-polling audits \citep{lindemanStark12,lindemanEtal12} also draw ballots at random
(uniformly, with replacement) and interpret the votes by hand, continuing to
draw ballots until there is strong evidence that the winners reported by the voting system
are the true winners.
Again, the strength of the evidence can be measured by $P$-values---one for each
(winner, loser) pair.
The $P$-values depend on the number of votes in the sample for each candidate.
Votes for a reported winner decreases all the $P$-values for pairs that include that winner;
a vote for a reported loser increases all the $P$-values for pairs that include that loser.
Hence, treating a ballot in the sample as if it simultaneously showed valid votes
for all the losers cannot decrease the $P$-value.
(The rules of the election might consider a ballot marked that way to
be an overvote, but that does not matter: We are talking about the effect
on the $P$-values, not how such a hypothetical ballot would be tabulated.)

Both types of audits involve drawing ballots at random with replacement from the ballots
cast in the contests under audit.
In practice, this is carried out by considering the groups in the ballot manifest
to be in some canonical order, considering the ballots in each group to be in some order,
and then conceptually numbering the ballots from 1 to the total number.
For instance, ballot~1 is the first ballot in the first group, ballot~2 is the second ballot
in the first group, and ballot~17,256 might be the 142nd ballot in the 89th group
(we call 142 the {\em index of the ballot\/} within the group).
We draw ballots at random by generating a random number between 1 and the total
number of reported ballots, then use the ballot manifest to figure out which ballot
corresponds to that number.
In this hypothetical, if we draw the random number 17,256, we audit the 142nd ballot in
the 89th group.
What happens if we go into the 89th group, and discover that it contains only 140~ballots?
Alternatively, what if we know that there should be 23,000 ballots in all, but the
manifest contains only 22,371?
What if we do not know how many ballots there should be, but we are sure there are no more
than 24,000, and the manifest lists only 22,371?
How can we conduct a rigorous risk-limiting audit in such cases?

It is perhaps obvious that if the errors in the manifest were known---for instance, if we knew
that exactly $24000 - 22371 = 629$ ballots were not listed in the manifest---assuming 
that there were an extra 629 votes for every losing candidate, adjusting the margins accordingly,
and auditing based on those adjusted margins would give a conservative procedure.
What is novel and (to us) surprising is that it is not necessary to adjust the margins to account
for those omissions.
Rather, it is enough to treat only the ballots that the audit attempts to find but cannot find
as votes for the losers (more generally, in the most pessimistic way)
Even more surprising is that it is not necessary to know how many ballots are missing, only
to have an upper bound. 

In sketch, \tooEz~is almost too easy:
If a ballot cannot be found (because the manifest is wrong---either because
it lists a ballot that is not there, or because it does not list all the ballots), pretend
that the audit actually finds a ballot, an evil zombie ballot that shows
whatever would increase the $P$-value the most.\footnote{%
   This is a related to the strategy \citet{benalohEtal11} use to account for possible
   errors in the cryptographic mapping from disaggregated votes to physical ballots.
  They assumed that the total number of ballots in the mapping was known to agree with the
  total number of physical ballots.
  Here, we do not necessarily assume that the total number of ballots according to
  the manifest is the total number of physical ballots.
}
For ballot-level comparison audits, this means pretending the ballot
would show a maximum overstatement of
2~votes; for ballot-polling audits, this means pretending it showed a valid vote for every loser
(even though the rules of the election might not allow that many valid votes on a single ballot).
While this clearly increases the $P$-value, the surprise is that it increases the $P$-value by
enough to ensure that the result is indeed conservative:
We prove that the $P$-value is stochastically larger than it would be if the 
manifest had been correct.
That is, we show that replacing ``missing'' ballots by hypothetical evil zombie ballots
compensates at least enough to make up for errors in the manifest.

We now consider the cases described above in increasing order of increasing complexity.

\section{Case~1.  $N_O$ is known}
Suppose we know $N_O$, the number of ballots actually cast, for instance, as the result of
ballot accounting and a compliance audit.
Let $N_M$ be the total number of ballots according to the manifest.

\subsection{$N_M = N_O$} \label{sec:NmEqNO}
If the total number of ballots according the the manifest equals the true total number,
then if some group has more ballots than the manifest claims, some other group must
have fewer than the manifest claims.
Suppose we sample a number uniformly from $\{1, \ldots, N_O\}$, then use the ballot
manifest to try to locate the corresponding ballot.
If a group has more ballots than the manifest claims, we will never sample it,
because the manifest maps numbers in the range corresponding to such ballots to
a different group.
Conversely, we might look for a ballot in a group and not find it, because its index
within the group exceeds the actual number of ballots in the group.
Suppose that, if that occurs, we pretend we actually found a ballot, and that
the ballot showed whatever would increase the $P$-value the most:
an overstatement of 2 votes for a ballot-level comparison audit, or
valid votes for all the losers for a ballot-polling audit.

We will call a group with more ballots than reported a {\em grave\/} and a group with
fewer ballots than reported {\em a hellmouth\/}.
In a grave, a ballot with an index larger than the number
of ballots the manifest claims the group has is {\em dead\/}:
The sampling scheme will never select it.
In a hellmouth, an index larger than the number of ballots in the
group represents a {\em phantom\/} ballot---one that, appears to be there
in the manifest but is not there in reality.

Imagine taking the dead ballots from all the graves and substituting them for phantoms,
putting just the right number in each hellmouth to make
the true number of ballots match the manifest in every group.
Then, the chance that the sampling scheme selects each ballot would be
what it should be: $1/N_O$ in each draw.
The dead ballots have been re-animated as zombies, replacing the phantoms
with real ballots.

Once the dead are re-animated as zombies, they become evil:
We suppose that they reflect whatever would increase the $P$-value 
most---a 2-vote overstatement for a ballot-level comparison audit,
or a valid vote for every loser in a ballot-polling audit.
Thus, if an evil zombie ballot is sampled, the $P$-value
will never be lower than was for the original votes on the dead ballot it came from.
Hence, the $P$-value is stochastically larger, and the chance of a full hand
count is higher: The risk-limit is even more conservative.

Our strategy of treating ballots that the manifest says should be there
but are not there when we look is mathematically equivalent to substituting
evil zombies for phantoms.
Hence, it is a conservative way to treat manifest error when
$N_O$ is the true number of ballots and $N_M = N_O$.

\subsection{$N_M < N_O$} \label{sec:NmLtNO}
Suppose $N_M < N_O$: There are fewer ballots in all according to the manifest than are known
to have been cast; the manifest omits some ballots.
If enough ballots were omitted from the manifest to alter the outcome (i.e., to
erase some margin), careful (possibly forensic) investigation and a full manual
check of the manifest, if not a full manual count of the votes, might be appropriate.

Otherwise, imagine adding a new group (a hellmouth) to the manifest that
contains the $N_O - N_M$ phantom ballots
that are missing from the manifest.
We now sample uniformly from $\{1, \ldots, N_O\}$.

If we draw a number greater than $N_M$, that corresponds to the new group that consists
of $N_O - N_M$ phantoms.
We don't know what vote that ballot might have shown, since we don't know where to find
it, but pretending that it would show a 2-vote overstatement (for a ballot-level
comparison audit) or valid votes for all the losers (for a ballot-polling audit)
decreases the $P$-values by the maximum possible.
That is, if we could find the ballot that was selected, the resulting $P$-value would be no larger
than it would be for this substitution.

If we draw a number less than or equal to $N_M$, we go into the appropriate group
and examine the ballot.
If that ballot turns out to be a phantom, again we again substitute a hypothetical
evil zombie ballot that ballot shows a 2-vote overstatement,
for a ballot-level comparison audit; or valid votes for all the losers, for a ballot-polling audit.
By the same argument as in section~\ref{sec:NmEqNO},
the resulting $P$-value is at least as large as it would be had the ballot actually been
where it was supposed to be according to the manifest.

Again, transforming phantom ballots into evil zombie ballots makes the
$P$-value stochastically larger than it would be for the correct manifest; hence,
the risk limit remains conservative.

\subsection{$N_M > N_O$} \label{sec:NmGtNO}
If the number of ballots according to the manifest exceeds the number known
to have been cast in the election,
this is prima facie evidence of ballot-box stuffing or some serious problem.
We suggest counting the groups of ballots again and examining the ballots
for evidence of fraud.
Careful, possibly forensic investigation and a full manual
check of the manifest might be appropriate.
We do not offer a statistical wooden stake.

\section{Case~2: An upper bound $N_U$ on $N_O$ is known} \label{sec:NuGeNO}
Suppose that we do not know the true number of ballots $N_O$, but we have an upper
bound $N_U$ on $N_O$.
If $N_U < N_M$, this is again prima facie evidence of ballot-box stuffing
or another serious problem, so---as in section~\ref{sec:NmGtNO}---we
suggest counting the groups of ballots again and inspecting them for
evidence of fraud.

If instead $N_U = N_M$ (the upper bound is equal to the number listed in the manifest),
then, unless the compliance audit has failed, $N_U = N_O = N_M$ and we
are in the case discussed in section~\ref{sec:NmEqNO}.

Now consider the case $N_U > N_M$.
The manifest {\em might\/} have omitted ballots, {\em possibly\/}
as many as $N_U - N_M$.
If $N_U - N_M$ ballots could alter the outcome,
again, careful investigation and a full manual
check of the manifest would seem appropriate.

Otherwise, as in section~\ref{sec:NmLtNO}, imagine augmenting the manifest with
a new group (hellmouth) of $N_U - N_M$ phantom ballots, ballots that might or might not
exist, but certainly are not listed in the manifest.
We sample uniformly from $\{1, \ldots, N_U\}$.
If this corresponds to a phantom ballot (either within one of the original groups or in the
new group), we replace it with an evil zombie: an overstatement of 2~votes
for a comparison audit, or valid votes for all the losers, for a ballot-polling audit.
This amounts to treating the $N_O - N_M$ actually missing ballots and
the $N_U - N_O$
nonexistent ballots as if they favored the losers.
In effect, this decreases the margin.
It increases the chance of selecting a ballot interpreted to increase the
$P$-value by as much as possible and
decreases the probability of selecting a ballot with every other
effect on the $P$-value.
This makes the $P$-value stochastically larger than it would be if we were sampling
from the actual ballots,
increasing the chance that the audit will lead to a full hand count.
Again, changing phantoms to evil zombies restores the risk limit, conservatively.

\section{Discussion}
Methods for ballot-level risk-limiting audits generally assume that it is possible
to sample ballots at random, uniformly.
To draw such samples, they rely on a ballot manifest, a description of how the
ballots are stored: a listing of the containers of ballots and the number of ballots
in each container.
If the manifest is wrong, the sampling distribution will be wrong.
If it is known that a certain number of ballots are missing from the manifest,
one could simply assume that that number of ballots would reflect votes
that would reduce all the margins as much as possible, adjust the margins
accordingly, and audit using the adjusted margins.
That approach would not help in situations where the number of ballots is
correct, but they are not organized as the manifest claims.
Moreover, it is unnecessarily conservative when the number of ballots listed in
the manifest is wrong.

Instead, we propose Phantoms to Evil Zombies (\tooEz).
\tooEz~samples uniformly, independently at random from
1~to the upper bound $N_U$ on the total number of ballots.
It uses the manifest to (try to) look up the corresponding ballot.
If, according to the manifest, the sampled ballot exists, but there is no such ballot
in the group, \tooEz~substitutes a ballot with the worst
effect on the $P$-value for this phantom ballot.
If, according to the manifest there is no ballot
corresponding to the selected ballot\footnote{%
   I.e., the random number is greater than $N_M$, the total number of
   ballots in the manifest.
}
\tooEz~substitutes a ballot with the worst effect on the $P$-value.
We prove that this takes into account errors and omissions in the ballot manifest in a way that
is guaranteed to be conservative:
The nominal risk limit is never smaller than the true risk limit.
If the manifest is in fact accurate and the upper bound on the number of ballots is tight,
\tooEz~has no effect on workload,
but if the manifest has errors or possible omissions, 
it requires the audit to look at more ballots---at least enough to attain the nominal risk limit.
Hence, we recommend that \tooEz~be used routinely with risk-limiting audits.

\bibliographystyle{apalike}
\bibliography{ref215proj}

\end{document}